\documentclass[12pt]{article}
\usepackage[utf8]{inputenc}
\usepackage[T1]{fontenc}
\usepackage{listings}
\usepackage{xcolor}
\usepackage[hmargin = 0.8 in, vmargin = 1in]{geometry}
\usepackage{amsmath}
\usepackage{amsthm}
\usepackage{amssymb}
\usepackage{caption}
\usepackage{esint}
\usepackage[parfill]{parskip}
\usepackage{listings}
\usepackage{floatrow}
\usepackage{graphicx}
\graphicspath{ {./Images-Lab10/} }
\usepackage{multirow}
\usepackage{multicol}
\usepackage{fancyhdr}
\usepackage{listings}    
\usepackage{hyperref}
\usepackage{caption}
\usepackage{subcaption}

\title{%
  CS754 - Advanced Image Processing  \\
  Sheel Shah - 19D070052\\
  Kushal Kejriwal - 190110040 \\
  Compressive Image Classification using Deterministic Sensing Matrices}
\author{}
\date{}

\begin{document}{

\newtheorem{theorem}{Theorem}
\newtheorem*{theorem*}{Theorem}
\newtheorem{corollary}{Corollary}[theorem]
\newtheorem{Lemma}[theorem]{Lemma}
\newtheorem{subLemma}[theorem]{Sub Lemma}
\newtheorem{defn}{Definition}[section]

\maketitle
\section*{Part 1 - Deterministic Sensing Matrices}
We have studied compressive sensing, where the goal is to capture majority of the \textit{information} in the signal using very few measurements and obtain accuracte reconstructions. One of the most important aspect of compressive sensing was the design of sensing matrices $\Phi$ and we understood that if the sensing matrices obey the Restricted Isometry Property, and obtained worst case performances on the reconstructed signals. 

Let the signal be $s$ sparse, let the dimension of the signal be $N$ and let the number of measurements be $m$. In particular, we looked at random sensing matrices that satisfy the RIP property with overwhelming probability with the overwhelming property given that:

\[ 
    m \geq \mathcal{O}(s\log(N/s))
\]

In our course, we have looked at random sensing matrices, and how they satisfy the RIP property that helps us in obtaining the \textbf{worst} case error bounds. This is analogous to Shannon's coding theory, that also provides a worst case error for a transmission channel and thus the random sensing matrices are associated with certain drawbacks which are described below:

\begin{enumerate}
    \item \textbf{Time Complexity during Reconstruction}: We have looked at a few algorithms to reconstruct signals from the measurements and some of them include the Basis Pursuit algorithm and the Orthogonal Matching Pursuit algorithm. The time complexity of these signals are given in Table \ref{tab:my_label}. In contrast, we can construct efficient reconstruction algorithms with lower time complexities. 
    \item \textbf{Storing the Random Sensing Matrix}: This requires significant space, especially if the dimension of signal is large. In contrast, entries of the deterministic sensing matrix can be computed on the fly, and thus no need of storage is required.
    \item \textbf{Verifying the RIP property}: There is no efficient algorithm for verifying whether the given sensing matrix obeys the RIP property. In the course we looked at an (inefficient) algorithm based on subset ennumeration with $\mathcal{O}(N^s)$ time complexity.
\end{enumerate}

\begin{table}[H]
    \centering
    \begin{tabular}{c|c}
        Algorithm & Time Complexity  \\
        \hline
         Basis Pursuit & $\mathcal{O}(N^3)$ \\
         Orthogonal Matching Pursuit & $\mathcal{O}(s^2\log^{\alpha}(N))$ \\
    \end{tabular}
    \caption{Time Complexities of Common Reconstruction Algorithms}
    \label{tab:my_label}
\end{table}

An important thing to consider here is analogous to coding theory, instead of the worst-case error we will analyse the typical (expected error) while studying deterministic sensing matrices. 

\subsection*{StRIP and UStRIP}
For dealing with deterministic sensing matrices, we will impose a weaker version of the Restricted Isometry Property, namely the Statistical Restricted Isometry Property.

\begin{defn}[$k, e, \delta$ - StRIP matrix]
An $m \times N$ sensing matrix $\Phi$ is said to be a $k, \epsilon, \delta $ - StRIP matrix, if for $k$ sparse vectors $\boldsymbol{\alpha} \in \mathbb{R}^N$, the following equation:
\[ 
    (1 - \epsilon)\lVert \boldsymbol{\alpha} \rVert^2 \leq \lVert \phi\boldsymbol{\alpha} \rVert^2 \leq (1 + \epsilon)\lVert \boldsymbol{\boldsymbol{\alpha}} \rVert^2
\]
holds with probability exceeding $1 - \delta$
\end{defn}

Note that this definition does not imply unique reconstruction, even with an exceedingly high probability. This is because the number of $s$-sparse vectors $\boldsymbol{\alpha} \in \mathbb{R}^N$ such that there is a different $s$-sparse $\boldsymbol{\beta} \in \mathbb{R}^N$ for which $\Phi\boldsymbol{\alpha} = \Phi\boldsymbol{\beta}$ being small is a much more strict condition than number of $s$-sparse vectors $\boldsymbol{\beta} \in \mathbb{R}^N$ and $\Phi\boldsymbol{\alpha} = \Phi\boldsymbol{\beta}$ be small. Note that the StRIP definition guaranteed the latter condition but not the former, and this is why we define the Unique Statistical Restricted Isometry Property as follows:

\begin{defn}[$k, e, \delta$ - UStRIP matrix]
An $m \times N$ sensing matrix $\Phi$ is said to be a $k, \epsilon, \delta $ - UStRIP matrix, if $\Phi$ is a ($k, e, \delta$ - StRIP matrix) and
\[ 
    \{\boldsymbol{\beta} \in \mathbb{R}^N, \Phi\boldsymbol{\alpha} = \Phi\boldsymbol{\beta}\} = \{\boldsymbol{\alpha}\}
\]
holds with probability exceeding $1 - \delta$
\end{defn}

\subsection*{StRIP-able Matrices}

We will now look at a few simple design rules, which are sufficient to guarantee that $\Phi$ is a UStRIP matrix, and these properties are satisfied by a large class of matrices.

\begin{defn}
An $m \times N$ - matrix $\Phi$ is said to be $\eta$ - StRIP-able, where $\eta$ satisfies $0 < \eta < 1$, if the following conditions are satisfied:
\begin{itemize}
    \item \textbf{St1}: The rows of $\Phi$ are orthogonal, and all the row sums are 0
    \item \textbf{St2}: The columns of $\Phi$ form a group under pointwise multiplication as follows:
    \[ \text{For all } j,j' \in \{1,2,....,N\}, \text{ there exists a } j'' \text{ such that } \phi_j\phi_{j'} = \phi_{j''}\]
    \item  \textbf{St3}: For all $j \in \{2,....,N\}$ 
    \[    |\sum_x \phi_j(x)|^2 \leq N^{2 - \eta}\]
\end{itemize}
\end{defn}

\subsection*{Main Result}
\begin{theorem}\label{theorem:thm1}
Suppose the $m \times N$ matrix $\Phi$ is $\eta$ StRIP-able, and suppose $k < 1 + (m - 1)\epsilon$ and $\eta > 1/2$. Then there exists a constant c such that, if $m \geq \left(c\dfrac{k\log(N)}{\epsilon^2}\right)^{1/\eta}$, then $\Phi$ is $(k, \epsilon, \delta) - UStRIP$ with $\delta = 2\exp\left(-\dfrac{[\epsilon - (k-1)/(N - 1)^2]m^{\eta}}{8k}\right)$
\end{theorem}

The authors also suggest the Quadratic Reconstruction Algorithm, which only requires vector-vector multiplication in the measurement domain, and thus has a sub-linear time complexity with respect to $N$ (the dimension of the signal). The StRIP property provides performance guarantees, because effectively the algorithm returns the location of one of the $k$ significant entries, and the property guarantees that the estimate is within $\epsilon$ of the true value. 

\subsection*{Delsarte - Goethals Frames}
Having introduced deterministic matrices for the purpose of compressive classification, we will now look at a particular type of deterministic matrices - the Delsarte-Goethals Frames. We will then show that these matrices are StRIP-able by proving that they obey the properties mentioned above. We will also show that these matrices when used for the purpose of compressive classification using Support Vector Machines give performance bounds with respect to classification in the data domain. 

\begin{defn}
A group is a set of elements $\Omega$ with some operation $*$ such that $*$ is associative, invokes an identity element in $\Omega$ and $\forall \ x, y   \in  \Omega, x*y \in \Omega$\\
\end{defn}

\begin{defn}
A Delsarte-Goethals set DG(m, r) is a set of $2^{(r+1)m}$ $m \times m$ binary symmetric matrices with the property that $\forall A, B \in DG(m, r)\ and\ A \neq B, rank(A - B) \geq m - 2r$\\
\end{defn}

\begin{defn}
A Delsarte-Goethals frame G(m, r) is a $2^m \times 2^{(r+2)m}$ matrix with each element defined as:
\[ a_{(P, b), t} = {i^{wt(d_P) + 2wt(b)}\cdot i^{tPt^T + 2bt^T}}/\sqrt{m}\]
where:
\[ t  \text{ is a binary m-tuple used to index rows of the matrix } \]
\[ P \text{ is a } m \times m \text{ matrix in } DG(m, r), \]
\[ b \text{ is a binary m-tuple } \]
\[ (P, b) \text{ indexes rows }\]
\[ wt(\cdot) \text{ is the Hamming weight of a binary vector (i.e.number of ones), } \]
\[ d_P \text{ is the diagonal of } P \]
\end{defn}

\begin{Lemma}\label{Lemma:dg1}
The set of unnormalized columns of G(m, r) denoted as $\Psi_{(P, b)}$ forms a group under point-wise multiplication ($*$).

\end{Lemma}
\begin{proof}
\begin{subLemma}
\begin{align*}
    For\ t \in  \mathbb{F}_2^m: \Psi_{(P, b)}(t) &= i^{wt(d_P) + 2wt(b)}\cdot i^{tPt^T + 2bt^T}\\
    \therefore \Psi_{(P, b)}(t)\Psi_{(P', b')}(t) &= i^{wt(d_P) + wt(d_{P'}) + 2wt(b) + 2wt(b')}\cdot\\
    & \ \ \ \ i^{tPt^T + tP't^T + 2bt^T + 2b't^T}\\
\end{align*}
\end{subLemma}

\begin{subLemma}
Let $\oplus$ denote binary addition, and construct $Q$ such that $P + P' = P \oplus P' + 2Q (mod 4)$. Hence, Q will be binary symmetric too, with $d_Q = dt* d_{P'}$. Moreover, it can be checked that $tQt^T = d_qt^T$, owing to Q being binary symmetric. This implies: 
\[ tPt^T + tP't^T = t(P+P')t^T = t(P \oplus P')t^T + 2tQt^T\\ =t(P \oplus P')t^T + 2(d_P * d_{P'})t^T\]
\end{subLemma}

\begin{subLemma}
$2x + 2y = 2(x \oplus y) mod 4$ and $2wt(x) + 2wt(y) = 2wt(x \oplus y)\ (mod 4)$. This can be checked by trying out all 4 possible binary combinations for a corresponding element of $x$ and $y$. This implies: 
\[ 2wt(b) + 2wt(b') = 2wt(b \oplus b') \]
\[ 2wt(b \oplus b') = 2wt(b \oplus b' \oplus 0) = 2wt(b \oplus b' \oplus d_P * d_{P'} \oplus d_P * d_{P'})\\= 2wt(b \oplus b' \oplus d_P * d_{P'}) + 2wt(d_P * d_{P'}) \]

\end{subLemma}

\begin{subLemma}
$wt(d_{P \oplus P'}) = wt(d_P) + wt(d_{P'}) + 2wt(d_P * d_{P'})$. This can be checked via taking cases for binary elements of $d_P, d_{P'}$.
\end{subLemma}

Combining all Sub Lemmas gives:
\begin{align*}
    \therefore \Psi_{(P, b)}(t)\Psi_{(P', b')}(t) &= i^{(wt(d_P) + wt(d_{P'}) + 2wt(d_P * d_{P'})) + 2wt(b \oplus b' \oplus d_P * d_{P'})}\cdot\\
    &\ \ \ \ i^{t(P \oplus P')t^T + 2(d_P * d_{P'})t^T + 2bt^T + 2b't^T}\\
    \therefore \Psi_{(P, b)}(t)\Psi_{(P', b')}(t) &= i^{(wt(d_P) + wt(d_{P'}) + 2wt(d_P * d_{P'})) + 2wt(b \oplus b' \oplus d_P * d_{P'})}\cdot\\
    &\ \ \ \ i^{t(P \oplus P')t^T + 2(d_P * d_{P'} \oplus b \oplus b')t^T}\\
    \therefore \Psi_{(P, b)}\Psi_{(P', b')} &= \Psi_{(P \oplus P', b \oplus b' \oplus d_P * d_{P'})}
\end{align*}
\end{proof}

\begin{Lemma}\label{Lemma:dg2}
G(m, r) is a tight frame with redundancy $\frac{n}{m}$. i.e. $G(m, r)G(m, r)^\dagger = \dfrac{nI_{m \times m}}{m}$\\
\end{Lemma}

\begin{proof}
We prove this showing that the inner product of two rows of $G(m, r)$ is $0$ if the rows are unequal and $\frac{n}{m}$ if they are equal.\\
Consider rows of index $t, t' \in \mathbb{F}_2^m$
\begin{align*}
    \therefore <G(m, r)[:, t], G(m, r)[:, t']> &= \sum_{P, b} \dfrac{\Psi_{P, b}(t)}{\sqrt{m}} \overline{\dfrac{\Psi_{P, b}(t')}{\sqrt{m}}}\\
    &= \frac{1}{m} \sum_{P, b} i^{tPt^T - t'Pt'^T + 2bt^T - 2bt'^T}\\
    &= \frac{1}{m} \sum_{P}  i^{tPt^T - t'Pt'^T} \sum_{b} i^{2bt^T - 2bt'^T}\\
    &= \frac{1}{m} \sum_{P}  i^{tPt^T - t'Pt'^T} \sum_{b} i^{2b(t \oplus t')^T}\\
    &= \frac{1}{m} \sum_{P}  i^{tPt^T - t'Pt'^T} \sum_{b} (-1)^{b(t \oplus t')^T}\\
    &= \frac{1}{m} \sum_{P}  i^{tPt^T - t'Pt'^T} \sum_{b' \in \mathbb{F}_2^{m'}} (-1)^{wt(b')}\\
    &\text{(where $m' = wt(t \oplus t')$)}\\
    &= \begin{cases}
    0\ if\ t \oplus t' \neq \textbf{0},\\
    \frac{n}{m}\ if\ t \oplus t' = \textbf{0}
    \end{cases}
\end{align*}
Notice that $m' = 0 \iff t \oplus t' = \textbf{0}$. Also, the pre-final step is justified because dot product with $(t \oplus t')$ is equivalent to adding those indexes of $b$ where $t \oplus t'$ is $1$, and since $b$ is varied over $\mathbb{F}_2^m$, this dot product is equivalent to the hamming weight of a binary sub-tuple of $b$ that selects the corresponding indices. Finally, it can easily shown that $\forall m' \geq 1, \mathbb{F}_2^{m'}$ has as many odd-hamming-weight vectors as even-hamming-weight ones, implying that the above sum is 0 for $m' \geq 1$.

\end{proof}

\begin{Lemma}\label{Lemma:dg3}
The column sums in the  $DG(m,r)$ satisfy:
\[ |\sum_{x}\phi_{P,b}(x)|^2 = 0 \text{ or } N^{2 - r/m}\]
\end{Lemma}
Now given Lemmas \ref{Lemma:dg1},\ref{Lemma:dg2} and \ref{Lemma:dg3}, we can conclude that $DG(m,r)$ frame is $r/m$ StRIP-able and thus obeys the UStRIP

\section*{Part 2 - Support Vector Machines for Compressive classification}

\subsection*{Classification using Support Vector Machines}
Before showing performance bounds on deterministic matrices, we will start off by first proving the error on performance bounds on matrices obeying the Restricted Isometry Property. Note that, random matrices obey the RIP property with overwhelming probability given that the number of measurments is large. The theorem can then be extended to deterministic matrices obeying the UStRIP. 

\subsubsection*{Notation}

\textbf{Data domain:} $\mathcal{X} = \{(\boldsymbol{x}, y): \boldsymbol{x} \in \mathbb{R}^n, y \in \{-1, 1\}, ||\boldsymbol{x}||_0 \leq k, ||\boldsymbol{x}||_2 \leq R\}$.\\

\textbf{Compressed sensing measurement matrix} $A \in \mathbb{R}^{m \times n}$.\\

\textbf{Measurement domain} $\mathcal{M} = \{(A\boldsymbol{x}, y): (\boldsymbol{x}, y) \in \mathcal{X}\}$.\\

Let the data be drawn from some unknown distribution $\mathcal{D}$ over $\mathcal{X}$. $S =\ <(x_1, y_1), ..., (x_M, y_M)>$ is a set of $M$ i.i.d samples from $\mathcal{D}$. $AS =\ <(Ax_1, y_1), ..., (Ax_M, y_M)>$.
Any linear classifier $w(x)$ corresponds to some $w \in \mathbb{R}^n$ such that $\boldsymbol{w(x)} = sign(\boldsymbol{w^Tx})$.\\

We will now define the loss functions as follows:
\begin{defn}
\begin{align*}
\text{The True Hinge loss }    H_{\mathcal{D}}(\boldsymbol{w}) &= E_{(\boldsymbol{x, y}) \sim \mathcal{D}} [1 - y\boldsymbol{w}^T\boldsymbol{x}]\\
\text{The Empirical Hinge loss }
    \hat{H}_{\mathcal{S}}(\boldsymbol{w}) &= E_{(\boldsymbol{x_i}, y_i) \sim S} [1 - y_i\boldsymbol{w}^T\boldsymbol{x}_i]\\
\text{The True Reguralisation loss }    
    L(\boldsymbol{w}) &= H_{\mathcal{D}}(\boldsymbol{\boldsymbol{w}}) + \dfrac{||\boldsymbol{w}||^2}{2C}\\
\text{The Emperical Reguralisation loss }        
    \hat{L}(\boldsymbol{w}) &= \hat{H}_{\mathcal{S}}(\boldsymbol{w}) + \dfrac{||\boldsymbol{w}||^2}{2C}\\
\end{align*}
\end{defn}

The $\boldsymbol{w}$ for soft margin SVM classifier that minimizes $\hat{L}(\boldsymbol{w})$ can be expressed as:
\begin{align*}
     \boldsymbol{w} &= \sum_{i = 1}^M \alpha_i y_i \boldsymbol{x_i}\\
    where\ &0 \leq \alpha_i \leq \frac{C}{M}\ \forall i\\
    and\ &||\boldsymbol{w}||^2 \leq C\\
\end{align*}
\clearpage
We now define the following classifiers:
\begin{defn}
\begin{align*}
    \boldsymbol{w^*} &= arg\min_w L(\boldsymbol{w}) \text{ (in data domain)}\\
    \boldsymbol{z^*} &= arg\min_z L(\boldsymbol{z}) \text{ (in measurement domain)}\\
    \boldsymbol{\hat{w}}_S &= arg\min_w \hat{L}(\boldsymbol{w}) \text{ (in data domain)}\\
    \boldsymbol{\hat{z}}_{AS} &= arg\min_z \hat{L}(\boldsymbol{z}) \text{ (in measurement domain)}\\
\end{align*}
\end{defn}

\subsection*{Main Result}
We prove that the SVM classifier trained on the measurement domain performs, with high probability, almost as well as the best classifier in the data domain. This is done by bounding the difference of losses between (a) the best data domain classifier and the data domain trained SVM, (b) the data domain SVM and its projection onto the measurement domain, and (c) this projection and the SVM trained directly on the measurement domain.\\

\begin{figure}[H]
    \centering
    \includegraphics[scale=0.7]{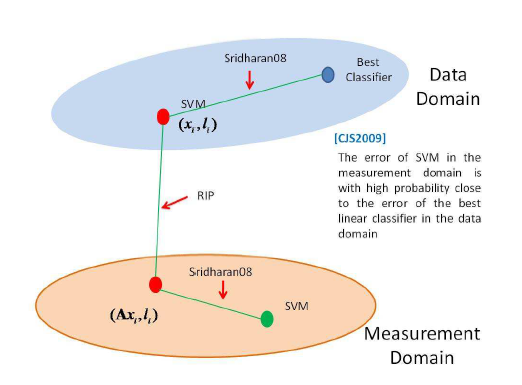}
    \caption{Hybrid Model for proving the performance bounds of the SVM classifier on the measurement domain - [3]}
    \label{fig:my_label}
\end{figure}

\begin{theorem}\label{theorem:thm2}
Let $A$ obey the RIP with $\delta_{2k} = \epsilon$, $AS, \hat{z}_{AS}$ be as defined above, and $w_0$ be the best data domain classifier. Then we have:
\[ H_D(\hat{\boldsymbol{z}}_{AS}) \leq H_D(\boldsymbol{w_0}) + \mathcal{O}\Bigg(\sqrt{||\boldsymbol{w_0}||^2\bigg(R^2 \epsilon + \frac{log(1/\delta)}{M}}\Bigg) \]

\end{theorem}

\begin{Lemma}\label{Lemma:lem10}
\[ (1 + \epsilon)\boldsymbol{x}^T\boldsymbol{x}' - 2R^2\epsilon \leq (A\boldsymbol{x})^T(A\boldsymbol{x}') \leq (1 - \epsilon)\boldsymbol{x}^T\boldsymbol{x}' + 2R^2\epsilon \]

\end{Lemma}

\ref{Lemma:lem10} essentially preserves the inner product in the measurement domain. This Lemma has been proved. We can now extend Lemma \ref{Lemma:lem10} to the inner product of linear combinations of sparse signals as follows.\\
\begin{Lemma}\label{Lemma:lem11}
Let $(\boldsymbol{x_1}, y_1), ..., (\boldsymbol{x_M}, y_M), (\boldsymbol{x_1}', y_1'), ..., (\boldsymbol{x_N}', y_N') \in \mathcal{X}$ and $\alpha_1, ..., \alpha_M, \beta_1, ..., \beta_N$ be non-negative numbers with $\sum_{i=1}^M \alpha_i \leq C$ and $\sum_{j=1}^N \beta_j \leq D$. Then for $\alpha = \sum_{i=1}^M \alpha_i y_i \boldsymbol{x_i}$, $\beta = \sum_{j=1}^N \beta_j y_j' \boldsymbol{x_j}' \leq D$, we have:
\[ |\boldsymbol{\beta}^T \boldsymbol{\alpha} - (A\boldsymbol{\beta})^T (A\boldsymbol{\alpha})| \leq 3CDR^2\epsilon \]
\end{Lemma}

The above result follows easily from the previously mentioned Lemma, and has been proved in Ref. 2

We will now prove one of the core results that connects the regularization loss of the data domain and the measurement domain
\begin{theorem}\label{theorem:thm3}
Let $\hat{\boldsymbol{w}}_S$ be as defined previously, and $A\hat{\boldsymbol{w}}_S$ be its projection. Then:
\[ L(A\hat{\boldsymbol{w}}_S) \leq L(\hat{\boldsymbol{w}}_S) + O(CR^2\epsilon) \]
Therefore, there exists at least one classifier in the measurement domain that doesn't perform much worse than the SVM classifier in the data domain.
\end{theorem}

\begin{proof}
We can write $\hat{\boldsymbol{w}}_S = \sum_{i = 1}^M \alpha_i y_i \boldsymbol{x_i}$, as seen above. Using Lemma \textcolor{red}{\ref{Lemma:lem11}}, with $N=M, D=C, (\boldsymbol{x_i}', y_i') = (\boldsymbol{x_i}, y_i)$, we get:
\begin{align*}
    (A\boldsymbol{w})^T(A\boldsymbol{w}) &\leq \boldsymbol{w}^T\boldsymbol{\boldsymbol{w}} + 3C^2R^2\epsilon\\
    \therefore \dfrac{||A\hat{\boldsymbol{w}}_S||^2}{2C} &\leq \dfrac{||\hat{\boldsymbol{\boldsymbol{w}}}_S||^2}{2C} + \mathcal{O}(CR^2\epsilon)\\
\end{align*}
Using Lemma \textcolor{red}{\ref{Lemma:lem11}}, with $N=M=1, D=C=1, (\boldsymbol{x_1}', y_1') = (\boldsymbol{x}, y)$, we get:
\begin{align*}
    1 - y(A\hat{\boldsymbol{w}}_S)^T(A\boldsymbol{x}) &\leq 1 - y\hat{\boldsymbol{w}}_S^T x + \mathcal{O}(CR^2\epsilon)\\
    \therefore H_D(-y(A\hat{\boldsymbol{w}}_S)^T(A\boldsymbol{x})) &\leq H(- y\hat{\boldsymbol{w}}_S^T \boldsymbol{x})+ \mathcal{O}(CR^2\epsilon)\\
\end{align*}

Since $A$ is one-one mapping (due to RIP), we can take expectation over $\mathcal{D}$ to complete the proof.
\end{proof}

As the pre-final step, we will now show that the regularisation loss of the SVM's classifier in the measurement domain ($\hat{\boldsymbol{z}}_{AS}$) is close to the projected regularisation loss $A\hat{\boldsymbol{w}}_S$. 
We will also show the regularisation loss of the SVM's classifier ($\hat{\boldsymbol{w}_S}$) in the data domain is close to the best classifier in the data domain ($\boldsymbol{w_0}$)

\begin{theorem}[\textbf{Sridharan 2008 [4]}]
For all $\boldsymbol{w}$ with $\lVert \boldsymbol{w} \rVert^2 \leq 2C$, with probability at least 1 - $\delta$ over training set:
\[
    L_{\mathcal{D}}(\boldsymbol{w}) - L_{\mathcal{D}}(\boldsymbol{w^*})\leq 2[\hat{L_{\mathcal{S}}}(\boldsymbol{w}) - \hat{L_{\mathcal{S}}}(\boldsymbol{w^*})]_+ + O\left(\dfrac{C\log(1/ \delta)}{M}\right)
\]
\end{theorem}

\begin{corollary}\label{cor:cor1}
Let $\hat{\boldsymbol{w}}_S$ be the SVM's classifier. Then with probability $1 - \delta$
\[ L_{\mathcal{D}}(\hat{\boldsymbol{w}}_S) \leq L_{\mathcal{D}}(\boldsymbol{w^*}) + O\left(\dfrac{C\log(1/ \delta)}{M}\right)\]
\end{corollary}

\begin{proof}
This directly follows from the fact that the SVM's classifier minimises the empirical regularisation loss, thus we have:
\[ \hat{L_{\mathcal{S}}}(\boldsymbol{w}) \leq \hat{L_{\mathcal{S}}}(\boldsymbol{w^*})\]

This can be now used to deduce the corollary.
\end{proof}

We are now ready to prove the main result of this section, i.e Theorem \ref{theorem:thm2}

\begin{proof}
By corollary \ref{cor:cor1}, we have a relation between the regularisation loss of the SVM's classifier in the measurement domain ($\hat{\boldsymbol{z}}_{AS}$) and the regularisation loss of the best classifier in the measurement domain ($\boldsymbol{z^*}$)
\[ L_{\mathcal{D}}(\hat{\boldsymbol{z}}_{AS}) \leq L_{\mathcal{D}}(\boldsymbol{z^*}) + O\left(\dfrac{C\log(1/ \delta)}{M}\right)\]

We also know that $\boldsymbol{z^*}$ is the best classifier in the measurement domain, thus we have:
\[ L_{\mathcal{D}}(\boldsymbol{z^*}) \leq L_{\mathcal{D}}(\boldsymbol{A\hat{w}_S})\]

We have bounded $L_{\mathcal{D}}(\boldsymbol{A\hat{w}_S})$ in Theorem \ref{theorem:thm3} by the regularization loss of the best classifier $\boldsymbol{w^*}$

\[ L_{\mathcal{D}}(\boldsymbol{A\hat{w}_S}) \leq L_{\mathcal{D}}(\boldsymbol{\hat{w}_S}) + O(CR^2\epsilon)\]
We can now apply the same corollary to the data domain to connect the the regularisation loss of the SVM's classifier in the data domain to the regularisation loss of the best classifier in the data domain ($\boldsymbol{w}^*$)
\[ L_{\mathcal{D}}(\hat{\boldsymbol{w}}_{AS}) \leq L_{\mathcal{D}}(\boldsymbol{w^*}) + O\left(\dfrac{C\log(1/ \delta)}{M}\right)\]

We can now complete the proof by noting that for any good classifier on $\mathcal{S}$ (say  $\boldsymbol{w_0}$), we have:
\[ L_{\mathcal{D}}(\boldsymbol{w^*}) \leq L_{\mathcal{D}}(\boldsymbol{w_0})\]
We can now combine all the inequalities to get:
\[ H_{\mathcal{D}}(\hat{\boldsymbol{z}}_{AS}) \leq H_{\mathcal{D}}(\boldsymbol{w_0}) + \dfrac{1}{2C}\lVert \boldsymbol{w_0} \rVert^2 + O\left(CR^2\epsilon + \dfrac{Clog(1/\delta)}{M}\right) \]

\end{proof}

\subsection*{Result for Deterministic Matrices}
We will now extend Theorem \ref{theorem:thm2} for deterministic matrices. Unfortunately, this theorem has just been mentioned in the paper but not proven by the authors. 
\begin{theorem}\label{theorem:thm4}
Let $o$ be an odd integer, and let $r \leq \dfrac{o-1}{2}$. Let A be an $m \times n$ DG frame with $m = 2^o$, and $n = 2^{(r+2)o}$. Let $\boldsymbol{w_0}$ denote the data-domain oracle classifier. Also let S = $\langle (\boldsymbol{x_1, l_1}), (\boldsymbol{x_M, l_M})\rangle$ represent $M$ training examples in the data-domain, and let AS = $\langle (\boldsymbol{y_1, l_1}), (\boldsymbol{y_M, l_M})\rangle$ denote the training examples in the measurement domain. Let $\hat{\boldsymbol{z}}_{AS}$ denote the measurement domain SVM classifier trained on S. Then there exists a universal constant $C$ such that if:
\[ m \geq \left( \dfrac{2^{r+1}C\log n}{\epsilon_1}\right)^2\] then with probability atleast $1 - \dfrac{6}{n}$, $H_{\mathcal{D}}(\boldsymbol{\hat{z}}_{AS}) - H_{\mathcal{D}}(\boldsymbol{\hat{w}}_{0})$ is \[ O\left(  R\lVert \boldsymbol{w_0}\rVert\sqrt{\left(
\dfrac{2^r(\log M + \log N)}{\sqrt{m}} + \sigma + \dfrac{((1+\epsilon_1)\log n)}{M}\right)}\right) \]
\end{theorem}
\subsection*{Results}
We have used the Brodatz Texture Database that contains 111 images of the textures - Horizontal, Vertical and None. We used a 50 \% train-test split and used an SVM classifier on the data domain (on the wavelet coefficients) and the compressed domain (randomly generated Gaussian matrices as sensing matrices)
\begin{figure}[H]
    \centering
    \includegraphics[scale=0.7]{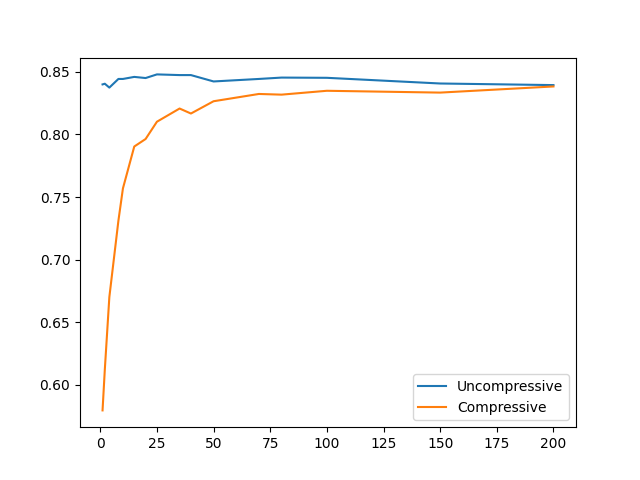}
    \caption{Training Accuracy vs Number of Measurements in Uncompressive and Compressive Sensing using an SVM classifier. We see that as $m$, the number of compressive measurements increases, the accuracy of SVM on the compressed data approaches that on the original, uncompressed data. Moreover, we see that for $m=200$, we have almost equal classification accuracy, implying a compression of $\frac{38*38}{200}$ = 7.22}
    \label{fig:my_label}
\end{figure}

\begin{figure}[H]
    \centering
    \includegraphics[scale=0.7]{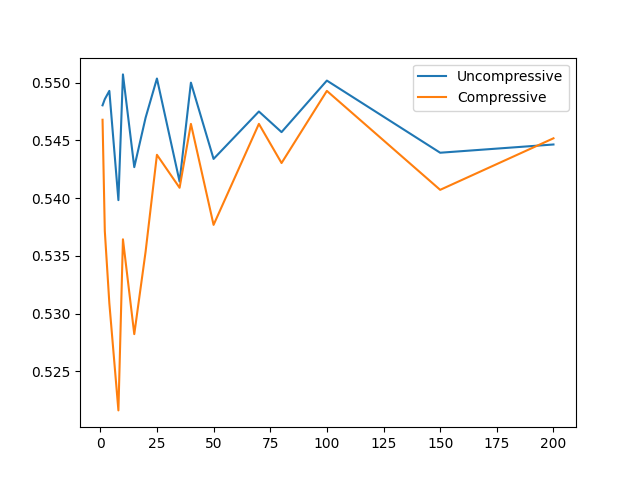}
    \caption{Test Accuracy vs Number of Measurements in Uncompressive and Compressive Sensing using an SVM classifier}
    \label{fig:my_label}
\end{figure}
\section*{Part 3 - An Attempt at Using Fischers Linear Discriminant on Image Classification}
FLD is a transform learning based image classification technique that learns a transformation vector $\boldsymbol{W}$ given a set of input points $\boldsymbol{x_i}$, and classifies the transformed points $\boldsymbol{y_i}$ given by:
        \[ \boldsymbol{y_i = w^tx_i}\]
        Let $\boldsymbol{m_i}$ denote the mean vector of each class and $\boldsymbol{m}$ denote the mean vector of the entire data, i.e:
        \[ \boldsymbol{m_i} = \sum_{x_j \in C_i} \boldsymbol{x_j}/n_i\]
        \[ \boldsymbol{m} = \sum_{i=1}^{c}\sum_{x_j \in C_i} \boldsymbol{x_j}/N\]

        We define the following three matrices:
        \begin{enumerate}
        
        \item \[ \boldsymbol{S_W} = \sum_{i=1}^{c}\sum_{x_j \in C_i}(\boldsymbol{x_j - m_i})(\boldsymbol{x_j - m_i})^T \]
        ($\boldsymbol{S_W}$ is the within class scatter matrix)
        \item \[ \boldsymbol{S_B} = \sum_{i=1}^{c}\sum_{x_j \in C_i}(\boldsymbol{m_i - m})(\boldsymbol{m_i - m})^T\]
        ($\boldsymbol{S_B}$ is the between class scatter matrix)
        \item \[ \boldsymbol{S_T} = \boldsymbol{S_B} + \boldsymbol{S_W} = \sum_{j = 1}^{N}(\boldsymbol{x_j - m})(\boldsymbol{x_j - m})^T\]
        ($\boldsymbol{S_T}$ is the between class scatter matrix)
        \end{enumerate}
        
         The optimisation function (Fischers criterion) which we maximise with respect to the transformation vector $\boldsymbol{w}$ is given by:
    \[ \boldsymbol{J(W) = \dfrac{|W^TS_BW|}{|W^TS_WW|}} \]

Essentially, a good $\boldsymbol{W}$ should transform the mean vectors of each class as far as possible \textbf{and, (this is also what sets apart FLD from PCA)} relative to the variances of the transformed vectors, the difference between mean vectors have to be large (to avoid conflation - overlap of points).

The Fischers criterion is a generalised eigenvalue problem, and the solution is a set of $c-1 \ (\boldsymbol{S_B} \text{ is a sum of rank 1 matrices})$ generalised eigenvectors of the following form:
\[ \boldsymbol{S_Bw_i} = \lambda_i\boldsymbol{S_Ww_i}\]
\[ \boldsymbol{S_W^{-1}S_Bw_i} = \boldsymbol{S_Ww_i}\]

We can see that by definition of $\boldsymbol{S_W} \in \mathbb{R}^{n \times n}$, its rank is at most $N - c$ (we subtract $c$, since the vectors are normalised with respect to their mean vector $\boldsymbol{m_i}$. This is results in $\boldsymbol{S_W}$ becoming singular and thus non-invertible.

 To circumvent this problem, we project the points $\boldsymbol{x_i}$ onto an $N-c$ dimensional space and then perform FLD to reduce the dimensionality to $c-1$
        Let the transformation learned by PCA be given by $W_{pca}$ and let the transformation learned by FLD be given by $W_{fld}$. The modified FLD criterion becomes:
        \[ \boldsymbol{W_{pca} = argmax_{\boldsymbol{W}} |\boldsymbol{W^TS_WW}|}\]
        
        \[ \boldsymbol{W_{fld} = argmax_{\boldsymbol{W}} \dfrac{|\boldsymbol{W^TW_{pca}^TS_BW_{pca}W}|}{|\boldsymbol{W^TW_{pca}^TS_WW_{pca}W}|}}\]

\subsection*{Results}

We were successfully able to implement the FLD (and PCA) algorithm on the training set (as shown in Figure \ref{fig:fig3}), but unfortunately we did not get the desired results on the test set. More specifically, as shown in Figure \ref{fig:fig4}, we did not get the desired clusters of points.  

\begin{figure}[H]
    \centering
    \includegraphics[scale=0.8]{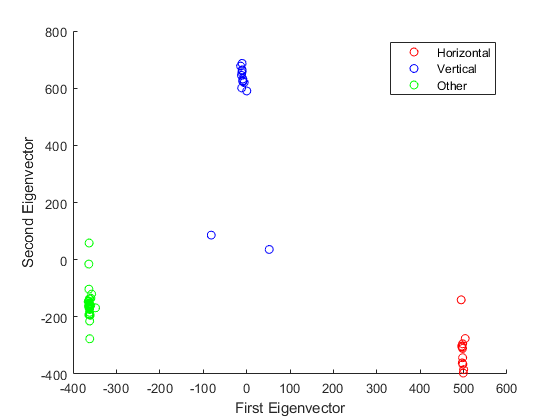}
    \caption{FLD (+ PCA) algorithm on the training set}
    \label{fig:fig3}
\end{figure}

\begin{figure}[H]
    \centering
    \includegraphics[scale=0.8]{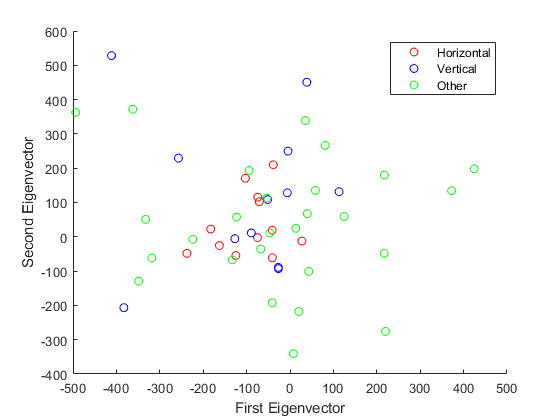}
    \caption{Training Accuracy vs Number of Measurements in Uncompressive and Compressive Sensing using an SVM classifier}
    \label{fig:fig4}
\end{figure}

\section*{References}
\begin{enumerate}
    \item \href{https://ieeexplore.ieee.org/document/6288656}{FINDING NEEDLES IN COMPRESSED HAYSTACKS}, \textit{Robert Calderbank and Sina Jafarpour}
    
    \item \href{https://ieeexplore.ieee.org/document/5419073}{Construction of a Large Class of Deterministic Sensing Matrices that Satisfy a Statistical Isometry Property} \textit{Robert Calderbank, Fellow, IEEE, Stephen Howard, Member, IEEE, and Sina Jafarpour, Student Member, IEEE}
    
    \item \href{http://citeseerx.ist.psu.edu/viewdoc/summary?doi=10.1.1.154.7564}{Compressed Learning:
Universal Sparse Dimensionality Reduction and Learning in the Measurement
Domain}
\textit{Robert Calderbank, Sina Jafarpour, Robert Schapire}

\item \href{http://ttic.uchicago.
edu/˜karthik/con.pdf, 2008.}{Fast
convergence rates for excess regularized risk with application
to SVM}\textit{K. Sridharan, S. Shalev-Shwartz, and N. Srebro. }

\item \href{https://cseweb.ucsd.edu/classes/wi14/cse152-a/fisherface-pami97.pdf}{Eigenfaces vs. Fisherfaces: Recognition
Using Class Specific Linear Projection} \textit{Peter N. Belhumeur, Joao~ P. Hespanha, and David J. Kriegman}
\end{enumerate}
}
\end{document}